\documentstyle[aps,pra,floats,epsfig]{revtex}

\begin{document}
\draft

\title{Measuring irreversible dynamics of a quantum harmonic oscillator}
\author{Giovanna Morigi,$^{1}$ Enrique Solano,$^{1,2}$ %
Berthold-Georg Englert,$^{1}$ and Herbert Walther$^{1}$}
\address{$^{1}$Max-Planck-Institut f{\"u}r Quantenoptik, %
Hans-Kopfermann-Strasse 1, 85748 Garching, Germany\\%
$^{2}$Secci\'{o}n F\'{\i}sica, Departamento de Ciencias, %
Pontificia Universidad Cat\'{o}lica del Per\'{u}, Apartado 1761, Lima, Peru}
\date{Rapid Communication submitted to Physical Review A,
received 17 August 2001}

\wideabs{
\maketitle

\begin{abstract}%
We show that the unitary evolution of a harmonic oscillator coupled 
to a two-level system can be undone by a suitable manipulation of 
the two-level system --- more specifically: by a quasi-instantaneous 
phase change. 
This enables us to isolate the dissipative evolution to which the 
oscillator may be exposed in addition. 
With this method we study the decoherence of a photon mode in cavity QED, 
and that of the quantized harmonic motion of trapped ions. 
We comment on the relation to spin echoes and multi-path interferometry.
\\{}PACS: 03.65.Yz, 32.80.Qk, 42.50.Ct\rule{0pt}{3ex}\end{abstract}}

\narrowtext

Decoherence is currently investigated theoretically and
experimentally~\cite{Decoherence}, thereby searching for the
boundary between the quantum and classical realms
\cite{Bertet01,Arndt99}. From the point of view of technical
applications, scalability of coherent control --- like the one
needed in a quantum computer --- is pursued intensively and
limited mainly by the presence of
decoherence~\cite{Decoherence,Bouwmeester00}. With the tools of
quantum optics, it has become possible to create quantum states of
high purity~\cite{Bouwmeester00} and to monitor their decay.

In particular, the decoherence of mesoscopic states of a quantized
electromagnetic field~\cite{Brune96} and of the quantized motion
of trapped ions~\cite{Myatt00} has been studied experimentally by
looking at the interference produced when recombining the two
classically distinguishable mesoscopic states. The decoherence of
{\it arbitrary} initial states of a quantum oscillator deserves
further studies, however.

In this contribution, we propose a method for measuring 
decoherence of a harmonic oscillator prepared in an arbitrary 
initial state by suitably coupling it to a stable two-level system, 
whose final level statistics carry the information of the oscillator decay.
The composite system evolves following a Jaynes-Cummings (JC)
interaction~\cite{Jaynes63} for a total duration $T$. 
At an intermediate instant, after time $\tau$ has been elapsed, the JC
evolution is interrupted by a phase kick, that is: a fast unitary
operation applied to the two-level system that introduces a
relative phase between the levels. 
We will show that, as a consequence of the phase kick, 
the coherent dynamics is effectively reversed, mimicking 
a time-reversal operation \cite{Vitali99}. 
When $\tau=T/2$, the final state of the composite system matches the
initial one, provided that the evolution is unitary. 
Any mismatch represents a measurement of changes in the oscillator 
state that are due to an external action.

This method allows to separate the irreversible decay of the
oscillator state from the unitary JC dynamics. 
The scheme can be easily implemented in two representative scenarios 
in Quantum Optics, ion traps and cavity QED, for studying the irreversible
decay of the quantized motion or the electromagnetic field,
respectively. 
In cavity QED, the harmonic oscillator is a mode of
the quantized radiation field, coupled to a resonant electronic
transition of atoms sent through the resonator and undergoing JC
dynamics~\cite{reviewCQED}. 
In ion traps, the center-of-mass motion of the trapped ion is harmonic, 
and it couples to an internal atomic transition when the ion is irradiated 
by a laser~\cite{reviewIons}. 
We discuss below how the decay of these
oscillators can be studied by means of the proposed scheme.

Let us consider the resonant interaction between a two-level
system, with lower state $|{\rm g}\rangle$ and upper state 
$|{\rm e}\rangle$, and a harmonic oscillator, with creation and
annihilation operators, $a^{\dagger}$ and $a$, respectively. 
The eigenstates of the number operator $a^{\dagger}a$ are denoted by
$|n\rangle$ with ${n=0,1,2,\ldots}$  
The transition energy of the two-level system is $\hbar\omega$, 
equal to the energy of the oscillator quanta. 
The coherent dynamics is described by the Hamiltonian
\begin{equation}
\label{Eq:1} H(t)=H_{\rm JC}+H_{\rm kick}=H_0+H_{\rm int}+H_{\rm kick}\,.
\end{equation}
Here, $H_{\rm JC}$ describes the JC dynamics, composed of the free
Hamiltonian
\begin{equation}
\label{Eq:2} H_0=\hbar\omega\left(\sigma^{\dagger}\sigma +
a^{\dagger}a\right)\,,
\end{equation}
with $\sigma=|{\rm g}\rangle \langle {\rm e}|$,
$\sigma^{\dagger}=|{\rm e}\rangle \langle {\rm g}|$, and of the
interaction term
\begin{equation}
\label{Eq:3} H_{\rm int}=\hbar g \left(\sigma^{\dagger}a + \sigma
a^{\dagger}\right)\,,
\end{equation}
where $g$ is the coupling constant. 
$H_{\rm kick}$ generates the instantaneous phase kick at $t=\tau$ 
and is defined as
\begin{equation}\label{eq:Hkick}
H_{\rm kick}=\hbar\pi\sigma\sigma^{\dagger}\delta(t-\tau)\,.
\end{equation}
The effect of the kick alone is thus given by
\begin{equation}\label{eq:kickop}
U_{\rm kick}={\rm e}^{-\frac{\rm i}{\hbar}\int_{\tau-0}^{\tau+0}{\rm d}t H(t)}
=\sigma^{\dagger}\sigma-\sigma\sigma^{\dagger}=\sigma_z \,.
\end{equation}
After the duration $T$, the evolution operator for the whole process is then
\begin{eqnarray}
U_{\tau}(T) &=&{\rm e}^{-\frac{\rm i}{\hbar}H_{\rm JC} (T-\tau)} U_{\rm kick}
{\rm e}^{-\frac{\rm i}{\hbar}H_{\rm JC} \tau}\nonumber\\
&=& \sigma_z {\rm e}^{-\frac{\rm i}{\hbar}(H_{0}
+\frac{2\tau-T}{T}H_{\rm int})T} \,, 
\label{Eq:4}
\end{eqnarray}
where the second line exploits $[H_0,H_{\rm int}]=0$ and
$\left\{H_{\rm int},\sigma_z\right\}=0$. 
So, the net evolution amounts to propagation governed by a JC coupling 
of the effective strength
\begin{equation}
\label{tune} g_{\rm eff}=\frac{2\tau - T}{T} g
\end{equation}
for time $T$, followed by the phase kick. 
Since the probability for finding the two-level system in 
$|{\rm g}\rangle$ or $|{\rm e}\rangle$ is measured eventually, 
this final phase kick can in fact be ignored. 
For $\tau=T/2$, the effective interaction term
in (\ref{Eq:4}) vanishes and the final state is just the freely
evolved initial state (except for the irrelevant phase kick). 
So, if the initial state were 
$|{\rm e}\rangle\otimes |\psi_{\rm osc}\rangle$, for example, 
the two-level system would always be in
the excited state after the interaction. 
It is as if the JC dynamics were reversed, which is somewhat reminiscent 
of spin-echo techniques~\cite{SpinEchoes}, where the evolution of
spin-$\frac{1}{2}$ particles in interaction with a {\em classical\/}
field is undone. 
We remark that, with this scheme, the external action could come from 
any other coherent or even incoherent process, but not from the probing 
two-level system. 
Nevertheless, it is its final statistics that carry the information about the
effect of the environment on the oscillator.

We now turn to the application of this scheme to the particular
scenario of cavity QED. 
The quantized electromagnetic field in a cavity is probed by atoms 
which fly through the resonator and whose dipole transition couples 
resonantly to a privileged field mode, undergoing the JC interaction 
described by $H_{\rm JC}$ in (\ref{Eq:1}). 
The phase kick could be implemented by a laser that couples
quasi-resonantly the ground state of the atom
to a third atomic level, thereby realizing a fast $2\pi$ pulse~\cite{ASW01}. 
We consider the proposed scheme for measuring the decay of the cavity field, 
coupled to a Markovian bath at zero temperature~\cite{reviewCQED}. 
We take for granted that the atomic transition is not subject to dissipation 
by itself on the relevant time-scale. 
The initial state $\rho(0)$ of the joint atom-field system is then evolved 
into the final state in accordance with
\begin{equation}
\rho(T) ={\rm e}^{L(T-\tau)}K{\rm e}^{L\tau}\rho(0)\,,
 \label{Full}
\end{equation}
where ${K\rho=U_{\rm kick}\rho U_{\rm kick}^{-1}=\sigma_z\rho\sigma_z}$ 
is the effect of the kick at time $\tau$, 
and the Liouville operator $L$ is given by
\begin{equation}
\label{L} L\rho=-\frac{\rm i}{\hbar}\left[H,\rho\right]+\kappa
\left( a\rho a^{\dagger} - \frac{1}{2}a^{\dagger}a\rho
-\frac{1}{2}\rho a^{\dagger}a\right)\,,
\end{equation}
where $\kappa$ is the field decay rate. 
This particular form of the non-unitary part of $L$ is chosen 
for convenient simplicity, but the method works just as well 
for other choices.

An explicit form of (\ref{Full}) can be obtained in
perturbation theory, in first order in the parameters 
${\kappa T\ll 1}$ and ${\kappa/g\ll 1}$. 
In this limit, assuming that the initial state is 
$\rho(0)=|{\rm g}\rangle\langle{\rm g}|\otimes%
\sum_{n,m}\rho_{nm}|n\rangle\langle m|$, 
the probability for measuring the atom at the exit 
in state $|{\rm g}\rangle$ is
\begin{eqnarray}
P_{\rm g} &=& 1-\sum_{n=2}^{\infty}\rho_{n,n}
\Bigl\{\frac{\kappa T}{4}(2n-1) \label{Result}\\
&&\mbox{}+\kappa\frac{\sin(gT\sqrt{n})}{4g\sqrt{n}}
-\kappa\frac{\sin(gT\sqrt{n-1})}{4g\sqrt{n-1}} \nonumber\\
&&\mbox{}-\frac{\kappa}{4g} \Bigl[\sqrt{n}(4n-3)\sin(gT\sqrt{n})
\cos(gT\sqrt{n-1})\nonumber\\
&&\mbox{}-\sqrt{n-1}(4n-1)\sin(gT\sqrt{n-1})
\cos(gT\sqrt{n})\Bigr]\Bigr\}\,. \nonumber
\end{eqnarray}
The effect of the decay is evident, causing deviations of $P_{\rm g}$ 
from unity. 
These deviations are oscillatory functions of the duration $T$, 
resulting from the JC dynamics characterizing the
system evolution. 
Note that in (\ref{Result}) the summation starts at $n=2$. 
In fact, for $\rho_{nn}=\delta_{n,1}$ one has $P_{\rm g}=1$ 
in this perturbative limit, the deviations appearing at
second order \cite{footnote}.

\begin{figure}[!t]
\centerline{\epsfig{file=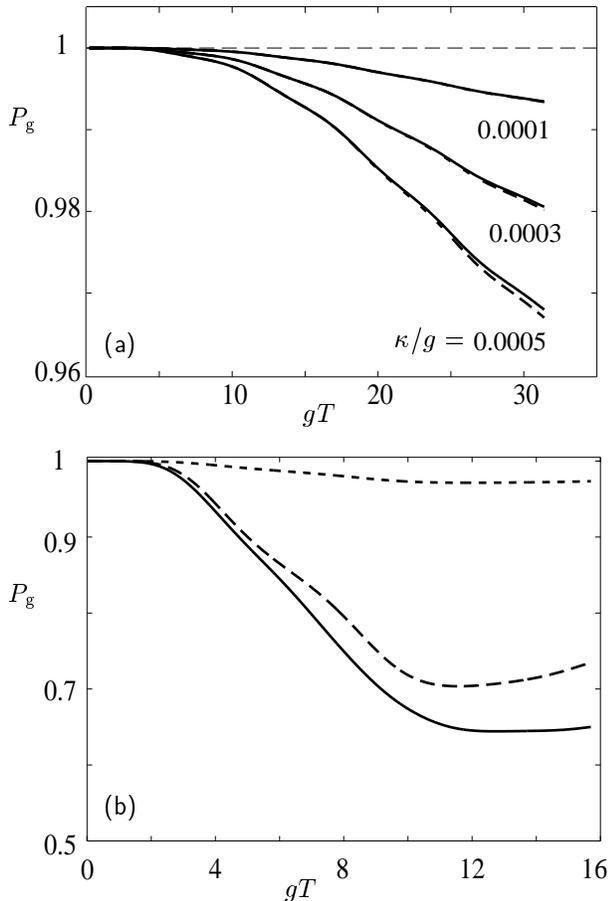}}
\caption{\label{Fig1}%
(a)~Numerical values [solid line, Eq.~(8)]
and approximate analytical values [dashed line, Eq.~(10)] of
$P_{\rm g}$ as a function of $gT$ for different ratios $\kappa/g$
and the field initially in the number state $|3\rangle$. 
(b)~Numerical values of $P_{\rm g}$ for $\kappa=0.05g$ and initial
cavity states $(|2\rangle+{\rm i}|3\rangle)/\sqrt{2}$ (solid
line), $|2\rangle$ (dashed line), coherent state $|\alpha\rangle$
with $\alpha=\exp({\rm i} \pi/4)/\sqrt{2}$ (dotted line).}
\end{figure}

In Fig.~\ref{Fig1}, we plot $P_{\rm g}$ as a function of the duration $T$.
Figure \ref{Fig1}(a) compares the full numerical evaluation of (\ref{Full}) 
with the approximation (\ref{Result}) for different decay 
rates $\kappa$. 
In Fig.~\ref{Fig1}(b), the curves are plotted for different initial
states. 
Note that for ${\kappa T\gg 1}$ the cavity state tends to
the vacuum, and the probability for counting the atoms in the
ground state approaches unity again.

\begin{figure}[!t]
\centerline{\epsfig{file=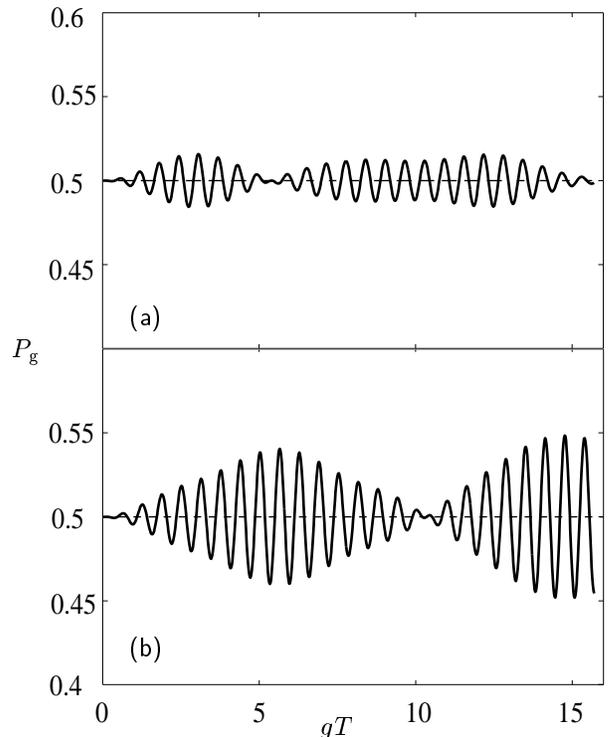}}
\caption{\label{Fig2}%
Phase sensitive $P_{\rm g}$ as a function of $gT$ for
$\kappa=0.05 g$ and $\omega=10g$. 
The initial field states are (a)~$(|2\rangle+{\rm i}|3\rangle)/\sqrt{2}$ 
and (b)~coherent state $|\alpha\rangle$ 
with $\alpha=\exp({\rm i}\pi/4)/\sqrt{2}$. 
In both plots, the constant $P_{\rm g}=0.5$ of an initial Fock state
is plotted for reference.} 
\end{figure}

The incoherent decay of the field phase can be studied by adding a
Ramsey zone at the exit of the cavity to effect an additional
unitary transformation afforded by 
$\exp\bigl(-{\rm i}\phi(\sigma{\rm e}^{{\rm i}\zeta} %
+ \sigma^{\dagger}{\rm e}^{-{\rm i}\zeta})\bigr)$. 
In this way, the Ramsey zone and the detectors constitute 
a field-phase sensitive detector~\cite{Englert96}. 
When only the coherent atom-field interaction takes place, 
we have $P_{\rm g}=\cos^2\phi$, independent of the initial state 
of the field, and the decay of the field phase is measured through 
the deviations from this value. 
In Fig.~\ref{Fig2}, we plot the phase sensitive probability
$P_{\rm g}$, after the Ramsey zone with $\phi=\pi/4$, $\zeta=0$,
as a function of the total interaction time for different initial
field states. 
The signals are modulated at the field frequency. 
For reference, in both figures we plot the curve corresponding to an
initial state characterized only by a diagonal density matrix
(dashed line) --- a Fock state, say.

We now study the application of the phase-kick scheme 
in ion traps, where a JC dynamics between the internal transition 
of the ions and its center-of-mass motion is implemented by a laser. 
Now, the Hamilton operator (\ref{Eq:1}) applies in
the reference frame rotating at the laser frequency. 
The atomic phase kick can be realized here in a similar way 
as in cavity QED case. 
Provided that the spontaneous emission from the atom can be
neglected and that the motion is coupled to a Markovian bath at
zero temperature, the evolution of the density matrix of the
composite system is again given by (\ref{Full}), and the results
reported above for the atom-cavity case are recovered. 
In the ion-trap situation, however, the phase kick is more easily 
implemented than in cavity QED. 
For example, we can switch off the laser system, after a fast initial 
coherent JC evolution and before the phase kick, and so interrupt the 
interaction during a convenient period of time $T_{\rm free}$. 
By continuing with the phase kick and a fast final coherent JC evolution, 
the experiment is completed.

This variation of our scheme has features in common with Ramsey
interferometry~\cite{Ramsey85}. 
One could  say that, during the initial short interaction time, 
interferometric paths are created in the ion-motion Hilbert space 
and that, after suffering ``dephasing'' during time $T_{\rm free}$, 
these paths are recombined into the initial $g_{\rm eff}=0$ superposition. 
With the aid of an arbitrary phase kick, instead of the fixed $2\pi$
kick, we create a modulated output that is reminiscent of an
interferometer pattern. 
This variation could also be implemented in the atom-cavity system, 
but the experimental setup would be much more demanding.

In general, the scheme can be applied to studying decoherence in
the collective motion of a chain of $N$ ions~\cite{Solano01},
i.e., a linear crystal of ions, achieved in linear ion
traps at sufficiently low temperature. 
In this limit, the chain motion along its main axis is described 
by the collective excitations of its $N$ normal modes with frequencies
${\nu_1,\ldots,\nu_N}$~\cite{James98}. 
A dynamics of the JC type can be designed between the $j$-th ion 
internal state and the $\beta$-th mode of the crystal, by laser 
irradiation analogous to the one-ion case. 
Then, the total Hamiltonian describing the interaction is given by 
$H_{\rm JC}^{(j,\beta)}=H_0^{(j,\beta)}+H_{\rm int}^{(j,\beta)}$ with
\begin{equation}
\label{Many}
H_0^{(j,\beta)}=\hbar\nu_{\beta}
\left(\sigma_j^{\dagger}\sigma_j^{\phantom{\dagger}} +
a_{\beta}^{\dagger}a_{\beta}^{\phantom{\dagger}}\right)\,,
\end{equation}
and
\begin{equation}
\label{Eq:Many} H_{\rm int}^{(j,\beta)}=\hbar g_{j,\beta}
\left(\sigma_j^{\dagger}a_{\beta}^{\phantom{\dagger}}
+ \sigma_j^{\phantom{\dagger}}a_{\beta}^{\dagger}\right)\,,
\end{equation}
where $g_{j,\beta}$ is the Rabi frequency for ion $j$ and mode
$\beta$, $a_{\beta}^{\phantom{\dagger}}$, $a_{\beta}^{\dagger}$ 
the annihilation and creation operators of a vibrational quantum 
$\hbar\nu_{\beta}$, while $\nu_{\beta}$ is the detuning of the laser 
from the atomic transition and equal to the frequency of the addressed mode. 
The dynamics, then, follows the lines of the one described for a
single ion, only that here the $\sigma_z$ kick is a laser pulse
applied to ion $j$. 
In this way, we are able to study the irreversible decay of a chosen 
collective mode. 
Similarly, it is possible to measure the irreversible loss of coherence of an
arbitrary motional state involving all modes. 
One would need to pair the ions with the modes, so that the corresponding
Hamiltonian has the form $H=\sum_{(j, \beta)} H_{\rm JC}^{(j,\beta)}$, 
a sum over these pairs. 
One could then measure the decoherence time of each mode, 
and possibly also of entangled states between different normal modes.

Here are some remarks on the experimental feasibility of the
proposed scheme in relation to the systems discussed before. 
First, we note that the phase kick is crucial for the whole scheme, 
but it is not necessary to realize the ideal kick of (\ref{eq:kickop}).
Indeed, the scheme is rather robust against experimental imperfections that
would amount to multiplying $H_{\rm kick}$ of (\ref{eq:Hkick}) by
${1+\epsilon}$. 
In particular, even with an error as large as $|\epsilon|\lesssim7\%$ 
we would still cancel more than 99\% of the JC dynamics.  

In the case of cavity QED, our scheme could be implemented in the
microwave and optical regime. 
For a test of the decoherence time associated with the population 
and the phase of the intracavity field, 
convenient ratios $\kappa/g\leq 1$, such as those presented
in Figs.\ \ref{Fig1} and \ref{Fig2}, can be found in both regimes. 
In either case, atomic decay is irrelevant on the time scale set by the
interaction. 
Furthermore, the kick could be implemented by introducing a sheet of 
light --- transversal to the mode of interest and crossed by the atomic 
trajectory at the center of the cavity --- by shining a laser through 
a hole drilled at the side of the cavity (closed cavity), 
or by focusing laser beams (open cavity). 
Note that, for measuring decoherence in atom-cavity systems, a quality
factor that is not too large may be desirable.
Further, the use of an external kick for testing this scheme is not 
necessary, if the field mode in question changes sign at the center 
of the cavity.
In this case, the odd spatial symmetry of the field undoes
naturally the JC dynamics, producing the same effect as the
phase kick for the fundamental mode.

Ion traps appear more fitting for testing and scaling the present
scheme. 
For example, the atomic decay can be neglected by suitably
coupling either a long-lived transition or two internal
metastable states with a two-photon coherent process. 
Also, JC (or anti-JC) dynamics like the one discussed here as well as 
pulses of well defined shape and duration are currently realized. 
Individual ions of a chain cannot be addressed by a laser in all experiments,
but this is not necessary for the present application. 
It is always possible to illuminate the chain of ions homogeneously and,
by giving an equally homogeneous phase kick, the JC dynamics could
be ``time-reversed'' and tested by measuring the global atomic
ground state. 
In this case, we replace $\sigma$ by ${\sum_j\sigma_j}$ in (\ref{Eq:Many}), 
and the phase-kick operator $\sigma_{z}$ by ${\prod_j\sigma_z^j}$. 

In summary, we have discussed a method for measuring the
decoherence of a harmonic oscillator by coupling it to a two-level
system and detecting its level statistics. 
The method is based on a phase kick applied to the two-level system 
during its interaction with the oscillator, thereby undoing the coherent
interaction and isolating the effect of any external disturbance
affecting the oscillator.
We have discussed applications of the scheme for measuring decoherence 
of the quantized field mode in cavity QED experiments and of the quantized 
motion of trapped atoms in harmonic traps. 
We think that the proposed scheme represents a useful tool for measuring 
decoherence times in different physical systems undergoing JC-like 
interactions. 
As a general feature, this method is sensitive to any deviation from
the expected JC dynamics and could be also used, for instance, for
measuring corrections to the rotating wave approximation, or for
studying the anharmonic coupling among the normal modes of a chain
of ions, or the influence of spectator atoms and spectator modes.

This work has been partly supported by the European
Commission (TMR networks ERB-FMRX-CT96-0077 and ERB-FMRX-CT96-0087).


\begin{references}
\bibitem{footnote}
For the initial state $\rho(0)=|{\rm g},1\rangle\langle {\rm g},1|$, 
one can evaluate $P_{\rm g}$ exactly,
\begin{displaymath}
P_{\rm g}=1-64 
\frac{g^2\kappa^2{\rm e}^{-\frac{\kappa T}{2}}}{(16g^2-\kappa^2)^2}
\sin^4\left(\frac{1}{8}\sqrt{16g^2-\kappa^2}T\right)\,,
\end{displaymath}
which shows that the effect of the decay is second order in
$\kappa/g$.
\end{references}
\end{document}